%% file: paper.tex
\documentclass[12pt]{aipproc}
\layoutstyle{6x9}
\usepackage{graphicx}
\usepackage{epsf}
\usepackage{epsfig}
\usepackage{color}

\let\bar=\overline

\begin{document}

\newcommand{\mb}{\mathbf}
\newcommand{\ssc}{\scriptscriptstyle}
\newcommand{\Plamc}{P_{\ssc\Lambda_{\ssc c}}}
\newcommand{\Lc}{\Lambda_c}
\newcommand{\Lcbar}{\overline{\Lambda_c}}
\newcommand{\hq}{\RedViolet{ c}}
\newcommand{\hqbar}{\RedViolet{\bar{c}}}
\newcommand{\vq}{\Blue{q}}
\newcommand{\vqbar}{\Blue{\bar{q}}}
\newcommand{\AR}{{{\cal A}_{\ssc R}}}
\newcommand{\Ps}{{{\cal P}_s}}
\newcommand{\PB}{{{\cal P}_{\ssc B}}}

\input mytex.tex
\input babar.tex
 \title{Production and Decay of the $\Lc$ Charmed
                       Baryon from Fermilab E791.}
 \author{B. Meadows
 \footnote{Representing the E791 Collaboration}}
 {address={University of Cincinnati, Cincinnati, OH, 45221, USA}}
 \begin{abstract}
 \input{abstract.tex}
 \end{abstract}
 \maketitle
 \input{production.tex}

 \input{decay.tex}

 \input{sample.tex}

 \input{asymm.tex}

\input{results.tex}

 \input{lamc_dk.tex}

 \input{summary}
 \bibliographystyle{$HOME/TeX/aipproc}
 \bibliography{the_bib}

\end{document}

%% file: mytex.tex
\newcommand{\ie}{{\sl i.e.~}}
\newcommand{\eg}{{\sl e.g.~}}
\newcommand{\etc}{{\sl etc~}}
\newcommand{\vs}{{\sl vs.~}}
\newcommand{\half}{{1\over 2}}
\newcommand{\z}{_{\circ}}
\newcommand{\pr}{^{\prime}}
\newcommand{\amu}{~{\rm u}}
\newcommand{\nm}{~{\rm nm}}
\newcommand{\fm}{~{\rm fm}}
\newcommand{\mps}{~{\rm m/s}}
\newcommand{\Hz}{~{\rm Hz}}
\newcommand{\eV}{~{\rm eV}}
\newcommand{\MeV}{~{\rm MeV}}
\newcommand{\eVcc}{~{\rm eV/c^2}}
\newcommand{\MeVcc}{~{\rm MeV/c^2}}
\newcommand{\kg}{~{\rm kg}}
\newcommand{\meter}{~{\rm m}}
\newcommand{\cm}{~{\rm cm}}
\newcommand{\J}{~{\rm J}}
\newcommand{\K}{~{\rm K}}
\newcommand{\W}{~{\rm W}}
\newcommand{\epsz}{\epsilon_{\circ}}
\newcommand{\muz}{\mu_{\circ}}
\newcommand{\grad}{\vec\nabla}
\newcommand{\curl}{\vec\nabla\times\vec}
\newcommand{\sidetxtfig}[2]{\leftline{%
\mbox{%
\begin{minipage}[b]{3.0in}{#2}%
\end{minipage}
\epsfig{file=#1,width=4.0in}
} 
} 
}%
\newcommand{\sidefigtxt}[2]{\leftline{%
\parbox{0.49\hsize}{#1}%
\hskip4mm
\parbox{0.49\hsize}{#2}%
} 
} 

%% file: babar.tex
\newcommand{\BaBar}{\mbox{$\mathrm{Ba\overline Bar}$}~}
\let\Babar=\BaBar
\let\mathrm=\rm
\newcommand{\PEP}{\mbox{$\mathrm{PEP}$}~}
\let\pep=\PEP
 
\newcommand{\hdr}[1]{{\Black{%
\small\phantom{.}\hfill#1\quad\today}}\newline\vspace{5mm}}
\newcommand{\sectn}[1]{\bigskip\noindent\underline{#1}\newline\noindent}

\newcommand{\CP}{CP\hskip-.5\em /}
\newcommand{\bz}{B^{\circ}}
\newcommand{\jpsi}{J/\psi}
\let\Jpsi=\jpsi
\newcommand{\KS}{K^{\circ}_s}
\newcommand{\KL}{K^{\circ}_L}
\newcommand{\Kz}{K^{\circ}}
\newcommand{\Kzbar}{\overline{K^{\circ}}}
\newcommand{\Kbar}{\overline{K}}
\newcommand{\Kst}{K^{\ast\circ}}
\let\Kstz=\Kst
\newcommand{\Kstbar}{\overline{K}^{\ast\circ}}
\let\Kstzbar=\Kstbar
\newcommand{\pz}{\pi^{\circ}}
\newcommand{\qbar}{\overline{q}}

\newcommand{\etal}{{\sl etal}~}
\newcommand{\wkph}{\delta_{\scriptscriptstyle W}}
\newcommand{\wkmix}{\delta_{\scriptscriptstyle M}}
\newcommand{\wkpp}{\wkph^{\prime}}
\newcommand{\wkpd}{\delta_{\scriptscriptstyle D}}
\newcommand{\wkpdp}{\wkpd^{\prime}}
\newcommand{\stph}{\delta_s}
\newcommand{\stpp}{\delta_s^{\prime}}
\newcommand{\Bz}{B^{\circ}}
\newcommand{\Bzbar}{\overline{B^{\circ}}}
\newcommand{\Bbar}{\overline{B}}
\newcommand{\Dz}{D^{\circ}}
\newcommand{\Dzbar}{\overline{D^{\circ}}}
\newcommand{\DMT}{{\Delta m\over 2}}
\newcommand{\fbar}{\overline{f}}


\newcommand{\bott}{\vfill\phantom{.}}

\newcommand{\rhoz}{\rho^{\circ}}
\newcommand{\etp}{\eta^{\prime}}
\newcommand{\mz}{m_{\circ}}
\newcommand{\fz}{f_{\circ}}
\newcommand{\ubar}{\overline{u}}
\newcommand{\dbar}{\overline{d}}
\newcommand{\sbar}{\overline{s}}
\newcommand{\bbar}{\overline{b}}
\newcommand{\cbar}{\overline{c}}
\newcommand{\tbar}{\overline{t}}
\newcommand{\pip}{\pi^+}
\newcommand{\pim}{\pi^-}
\newcommand{\piz}{\pi^{\circ}}
\newcommand{\Kp}{K^+}
\newcommand{\Km}{K^-}
\newcommand{\Dsp}{D_s^+}
\newcommand{\Dp}{D^+}
\newcommand{\Dst}{D^{\ast}}
\newcommand{\Dpst}{D^{\ast+}}
\newcommand{\Xcp}{\Xi_c^+}
\newcommand{\Xcz}{\Xi_c^{\circ}}

\newcommand{\Aj}{{\cal A_{\rm j}}}
\newcommand{\llik}{{\cal L}}
\newcommand{\aj}{\Red{a_{\rm j}}}
\newcommand{\deltj}{\Red{\delta_j}}
\newcommand{\fj}{\Red{f_{\rm j}}}
\newcommand{\bi}{\Red{b_{\rm i}}}

%% file: abstract.tex
Results are presented for the 500 GeV/c pion production asymmetry and
polarization of the $\Lc~(\Lcbar)$
charmed baryon from Fermilab experiment E791.
An analysis of the decay to the $p\bar{K}\pi$ final state is described.
Resonant sub-channel fractions {\bf and phases} are given and possible
resonant effects in the low mass $p\bar{K}$ system discussed.  
Significant decay to $\Lc\to\Delta^{++}K^-$ establishes for the first
time the importance of a $W$ exchange mechanism in charmed baryon decay.

%% file: production.tex
\newcommand{\hadr}{{\scriptstyle\cal P}}

Measurements of asymmetry
 $ A_{\cal P} =
   \left(d\sigma_{\hadr} - d\sigma_{\bar{\hadr}}\right) /
   \left(d\sigma_{\hadr} + d\sigma_{\bar{\hadr}}\right) $
 in the yield of particle $\cal P$ and anti particle $\bar{\cal P}$
 can provide information on the production mechanisms involved.
Dependences of $A$ on $x_F$ and $p_T^2$ can distinguish different
 production models.
Several experiments \citep[and refs 1-9 therein]{ds-asymmetries}
 have shown that production of
 charmed mesons is characterized by leading particle
 effects and that asymmetries can be large.
Leading particle behaviour has also been observed in production
 of strange hyperons in E791
 \citep{e791HyperonAsymmetries} - even in a
 very central region.

%% file: decay.tex
Branching fractions for baryon decays provide
 information on the relative importance of lowest order
 decay mechanisms - $W$ exchange or spectator processes.
In $\Lc\to pK\pi$ decay,
\footnote{Note that charged conjugate states are implied unless
stated otherwise.}
 $W$ exchange can contribute to $p\Kst$,
 $\Lambda^{\ast}\pi$, $\Sigma^{\ast}\pi$ or $pK\pi$
 channels, but for the $\Delta^{++}\Km$ mode it is the {\sl only} low
 order process possible.
Evidence for this decay requires a large sample of $pK\pi$ decays and
 proper analysis of interference effects in the system.

Reported here are the first published
 \citep{e791LambdaCAsymmetries}
 measurements of both $x_F$ and $p_T^2$ dependence of $A$ for charmed
 baryon production.  We also present \citep{e791LambdaCResonance} the
 first full analysis of charmed baryon decay, measuring
 $\Lc$ branching fractions, relative phases and polarization.

%% file: sample.tex

This study is based on a sample of $2\!\times\!10^{10}$ events produced
from the interaction of 500 GeV/c $\pi^-$ incident on thin foils,
one $Pt$ and four $C$.
$Pt$ target data (unequal numbers of $n$ and $p$) were not
used in the asymmetry study.
The detector and data reconstruction are described in
\citep{Aitala:1998kh}.
Cuts on geometric and kinematic quantities were made to identify
 $\Lc\to p\Km\pip$ decays.
The decay vertex had to be well separated ($>5\sigma$) from both
 production vertex and nearest target material.
\begin{figure}
 \label{fig-lamc_production}
 \begin{minipage}[ht]{0.54\textwidth}
 \centerline{%
 \epsfig{file=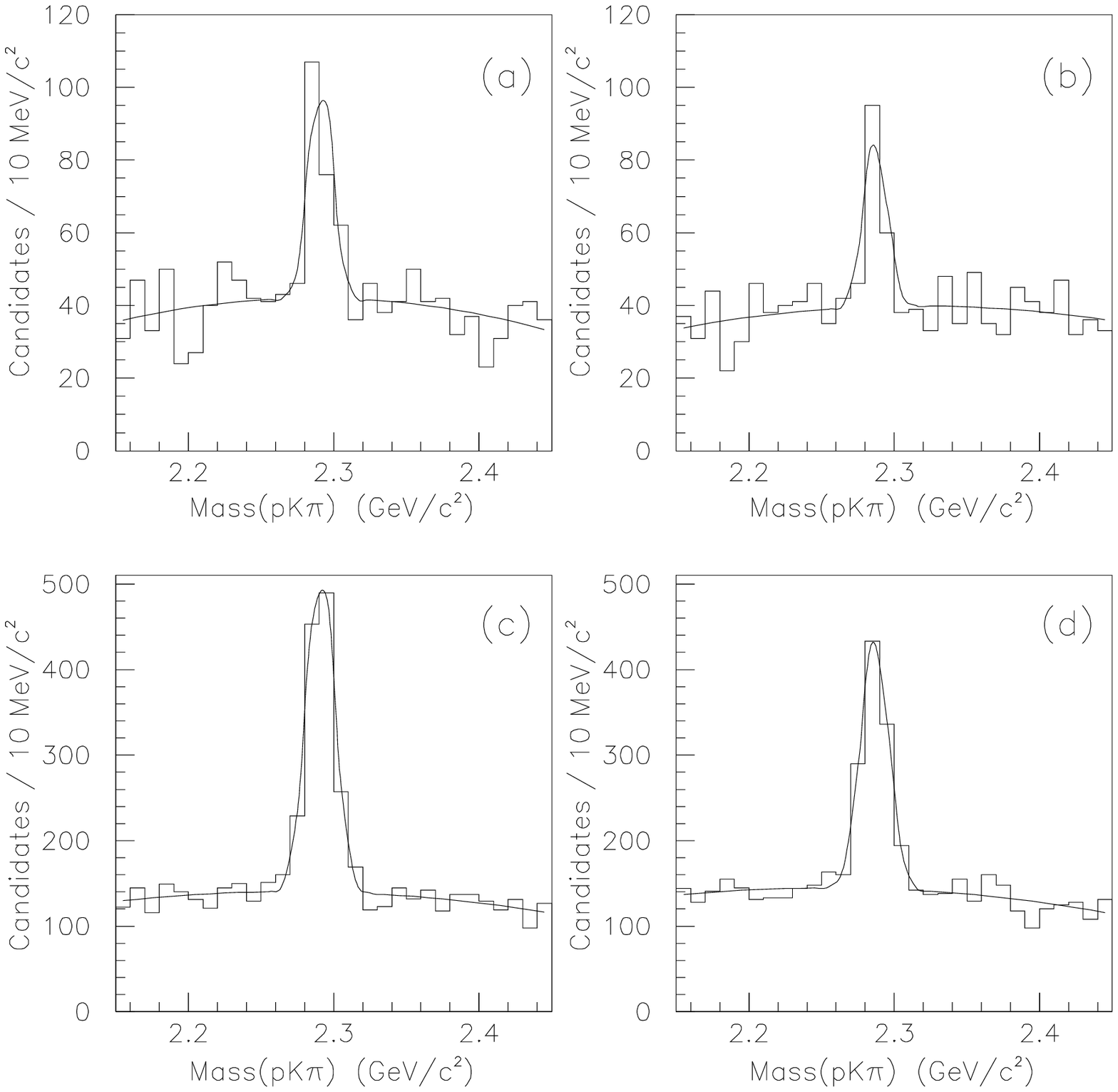,
         height=3.00in,width=3.24in,angle=0}}
 \par\vspace{0pt}
 \end{minipage}
 \begin{minipage}[ht]{0.44\textwidth}
 \centerline{%
 \epsfig{file=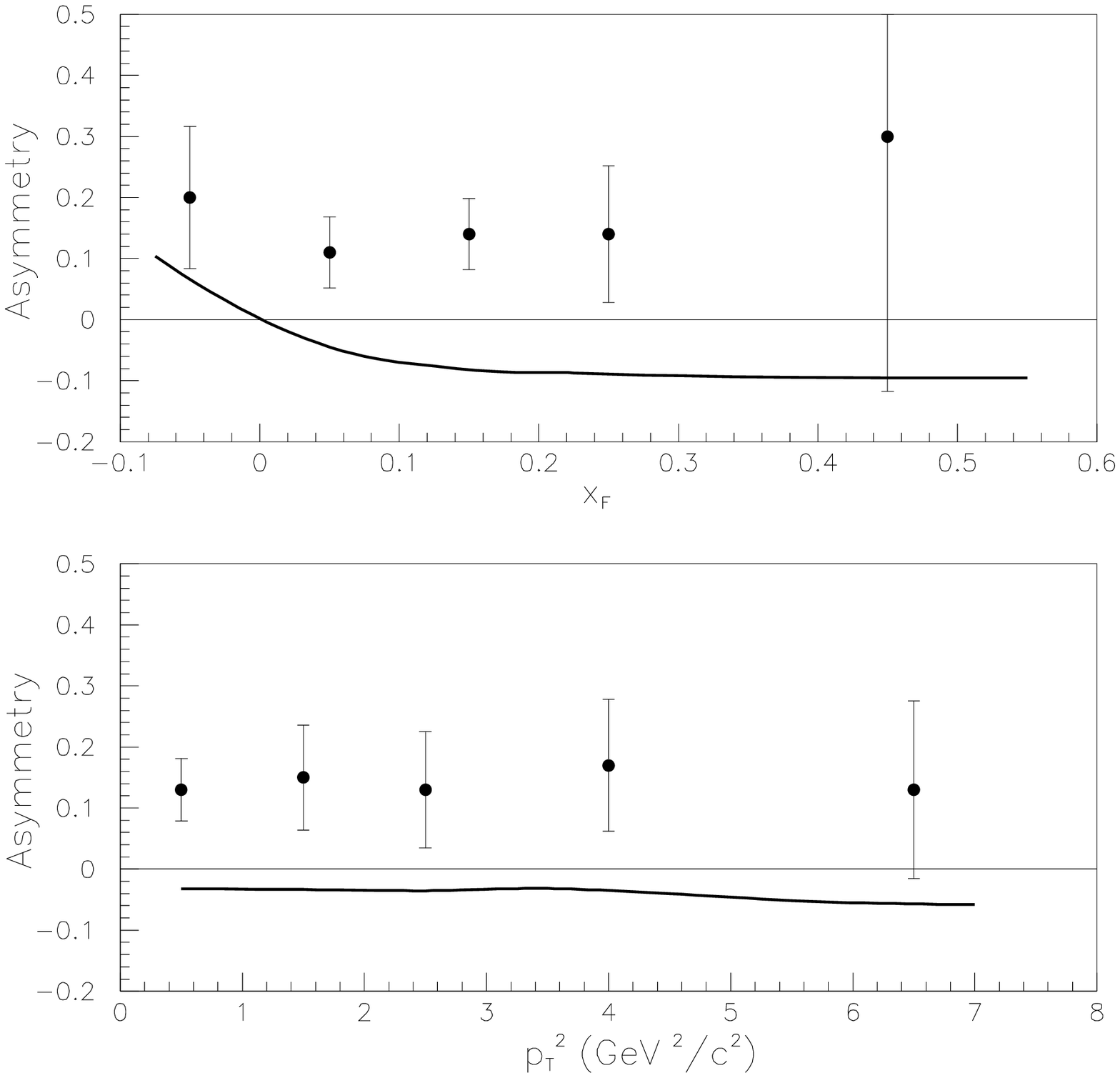,
         height=3.00in,width=2.64in,angle=0}}
 \par\vspace{0pt}
 \end{minipage}
 \\
 \caption{$\Lc (\Lcbar)$ samples: (a) $\Lc$ $(x_F<0)$;
 (b) $\Lcbar$ $(x_F<0)$;
 (c) $\Lc$ $(x_F>0)$;
 (d) $\Lcbar$ $(x_F>0)$.
 Fits shown in (a)-(d) are Gaussian peaks on 2nd order polynomial backgrounds.
 Asymmetries: (e) vs $x_F$  (f) vs. $p_T^2$.
 The solid curves in (e) and (f) are the prediction of Pythia/Jetset.}
\end{figure}
The yield, shown in Fig. \ref{fig-lamc_production}(a)-(d), was
 $ 1,025\pm 45~\Lc\!\to\!pK^-\pi^+$ and
 $794  \pm 42~\Lcbar\!\to\!\bar{p}K^+\pi^-$.
Events were divided into 5 regions of $x_F$ and 5 of $p_T^2$
 in the overall ranges:
 $ -0.1<x_F<0.6$ and $p_T^2\le 8~(GeV/c)^2 $
 chosen to have clear $\Lc$ signals in each.  Fits
 similar to those shown in the figure were made to each sample to
 determine the number $N (\Lc)$ and $\bar N(\Lcbar)$ of signal events
 in each range.

%% file: asymm.tex
\section{\ensuremath{\Lc/\Lcbar} Asymmetries}

Efficiencies $\epsilon~(\bar\epsilon)$ for $\Lc(\Lcbar)$ were
 not quite equal due to the asymmetric effect of the intense $\pim$ beam
 on the drift chambers.  This effect was greatest at large $x_F$ and low
 $p_T^2$.  It was necessary therefore to estimate the ratio
 $r=\epsilon/\bar\epsilon$ in each of the 5 $x_F$ and 5 $p_T^2$ ranges
 using Monte Carlo samples of $\Lc$ \& $\Lcbar$
 generated with Pythia/Jetset, projected through a simulated E791
 detector and subjected to the same reconstruction code and selection
 criteria as the data.
Corrected asymmetries $A~=~ (N-\bar{N}/r)/(N+\bar{N}/r)$ were then
 obtained in each range.

%% file: results.tex
The main sources of systematic uncertainty
 (parametrization of signal and background shapes and
 precision of $r$) amounted to less than 50\% of the
 statistical uncertainty in all instances.
The results are shown in Figure \ref{fig-lamc_production}(e) and (f) and
 compared with
 earlier $\pim N$ studies in Table \ref{tab-compare}.
\begin{table}[hbt]
 \label{tab-compare}
 \caption{Comparison with asymmetries (\%) from earlier
 $\pi^-N$ experiments.}
 \begin{tabular*}{0.8\textwidth}{ c c c c }
 $ x_F$ region~~~
               & E791~~~~~       
               & ACCMOR \citep{accmor}~~~
               &     SELEX \citep{selex-onepoint}~~~   \\
 \hline
 $ x_F<0$      &  $20 \pm 10  \pm 6$    &    $ -- $     &   $ -- $    \\
 $ x_F>0$      & $12.3 \pm 3.7 \pm 1.6$ & $0.5 \pm 7.9$ & $25 \pm 15$ \\
 \hline
 \end{tabular*}
\end{table}
%

The asymmetry is positive and flat throughout the range.  This
 might result from the additional energy required to produce additional
 baryons when a
 $\Lcbar$ is produced, favouring $\Lc$ production in general.
The solid curve in Figure \ref{fig-lamc_production} is the prediction of
 Pythia/Jetset and clearly does not describe the data well.
Two component intrinsic charm/coalescence models \citep{Vogt:1996fs},
 \citep{Herrera:1998qh} predict a rising asymmetry beginning at the low
 end or possibly below the range of this data.
Leading particle effects would also result in a rising
 asymmetry in the entire $x_F<0$ region.
The data do not rule that possibility out.

%% file: lamc_dk.tex
\newcommand{\lamp}{\pm\half}
\section{Analysis of the Decay \mbox{$\textstyle{\Lc\to p\Km\pip}$}}

A cleaner sample was required for this analysis.
The length cut was increased to $8\sigma$ and a neural net criterion
 was used to optimize the significance $S/\sqrt{S+B}$
 of the signal ($S=886\pm 43$) over background ($B\sim300$) in the
 fit region.

These decays were defined by five independent variables,
 \eg two Dalitz plot coordinates and orientation of the decay plane
 relative to production plane ($z$ axis).
The $\Lc$ could have polarization $\Plamc\hat z$.

Each isobar decay channel $\Lc\!\to\!R(\to\!ab)c$ was assigned
 an amplitude labeled by the $z$ component of $\Lc$ spin, $m$,
 and proton helicity ($\lamp$) in the $\Lc$ rest frame \\
 ${\cal A}^{\ssc R}_{m,\lamp}=B^{\ssc R}(M_{ab})
      (a_{\pm}e^{i\alpha_{\pm}}|m,\lamp,\lambda_{\alpha_{\pm}}> +
       b_{\pm}e^{i\beta_{\pm}}|m,\lamp,\lambda_{\beta_{\pm}}>)$
 where $\lambda_{\alpha_{\pm}}$ \& $\lambda_{\beta_{\pm}}$ are the two
 possible helicities for $R$ with unknown coefficients
 $a_{\pm}e^{i\alpha_{\pm}}$ \& $b_{\pm}e^{i\beta_{\pm}}$.
The $B^{\ssc R}(M_{ab})$ were Breit Wigner functions. For
 non-resonant $NR$ decay to $pK\pi$ a similar amplitude with $B=1$ was
 used.
An unbinned, maximum likelihood fit was used to determine
 $a_{\pm},\alpha_{\pm},b_{\pm},\beta_{\pm}$
 and three values for $\Plamc$ (one in each of three $x_F$ ranges).
The signal probability density function was
 \begin{eqnarray*}
   \Ps &=&\half\times\epsilon\times\mbox{[}
    (1+\Plamc)\left(
    \left|\sum_{\ssc R}{\cal A}^{\ssc R}_{ \half, \half}\right|^2~+~
    \left|\sum_{\ssc R}{\cal A}^{\ssc R}_{ \half,-\half}\right|^2\right)
    \\
       & &\mbox{} +
    (1-\Plamc)\left(
    \left|\sum_{\ssc R}{\cal A}^{\ssc R}_{-\half, \half}\right|^2~+~
    \left|\sum_{\ssc R}{\cal A}^{\ssc R}_{-\half,-\half}\right|^2\right)
    \mbox{]}
 \end{eqnarray*}
Five dimensional efficiency ($\epsilon$) and background density
 were estimated empirically from a MC sample and $M_{pK\pi}$ sidebands.
Modes included were
 $p\Kstbar(890)$,
 $\Delta^{++}(1232)\Km$,
 $\Lambda(1520)\pip$ and
 $NR$.

%
 \begin{figure}
 \label{fig-fit_results}
 \begin{minipage}[ht]{0.48\textwidth}
   \epsfxsize=2.8in
   \epsfysize=2.6in
   \centerline{\epsffile{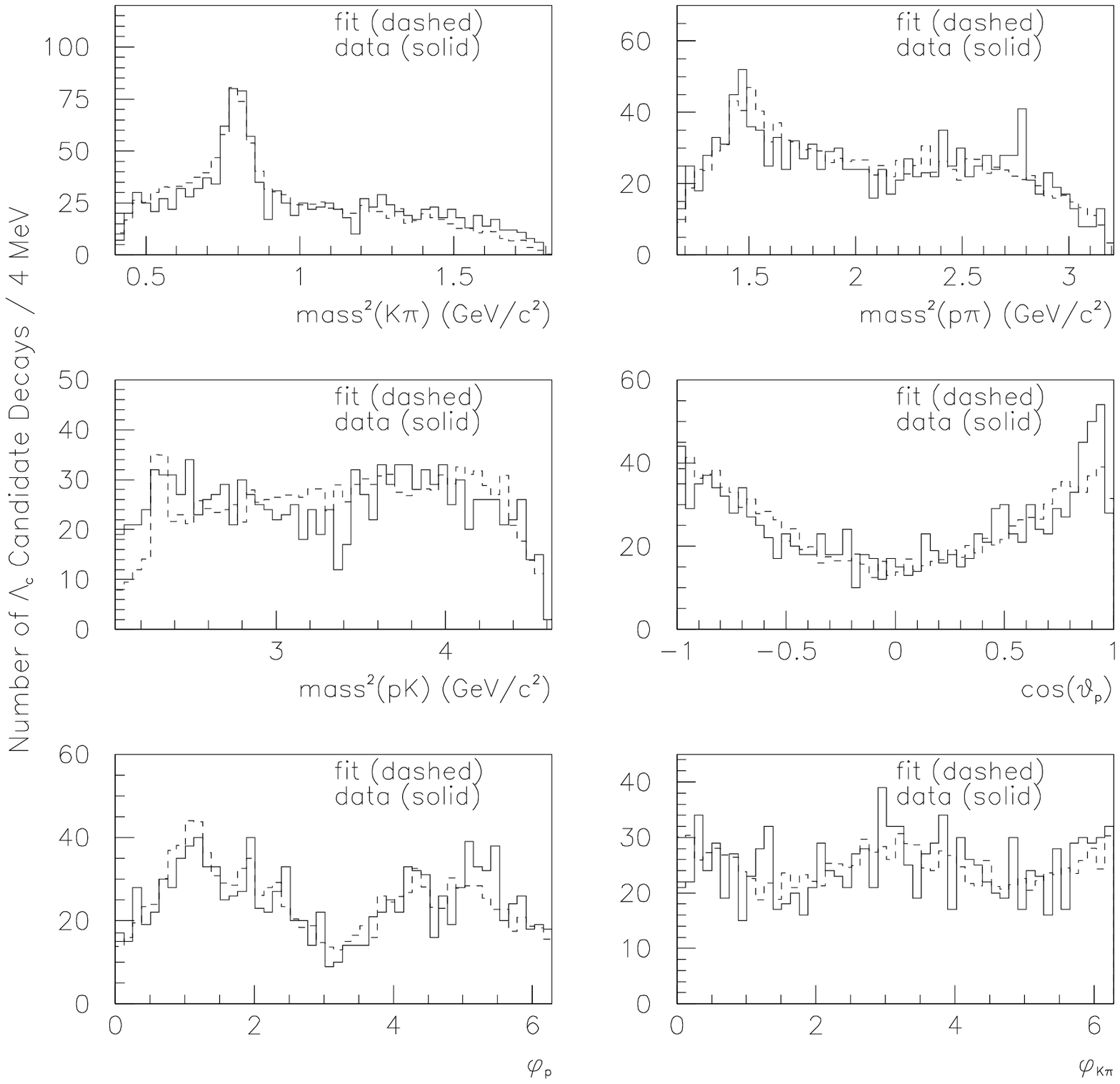}}
  \par\vspace{0pt}
 \end{minipage}
 \hskip0.04\textwidth
 \begin{minipage}[ht]{0.48\textwidth}
   \epsfxsize=2.8in
   \epsfysize=2.6in
   \centerline{\epsffile{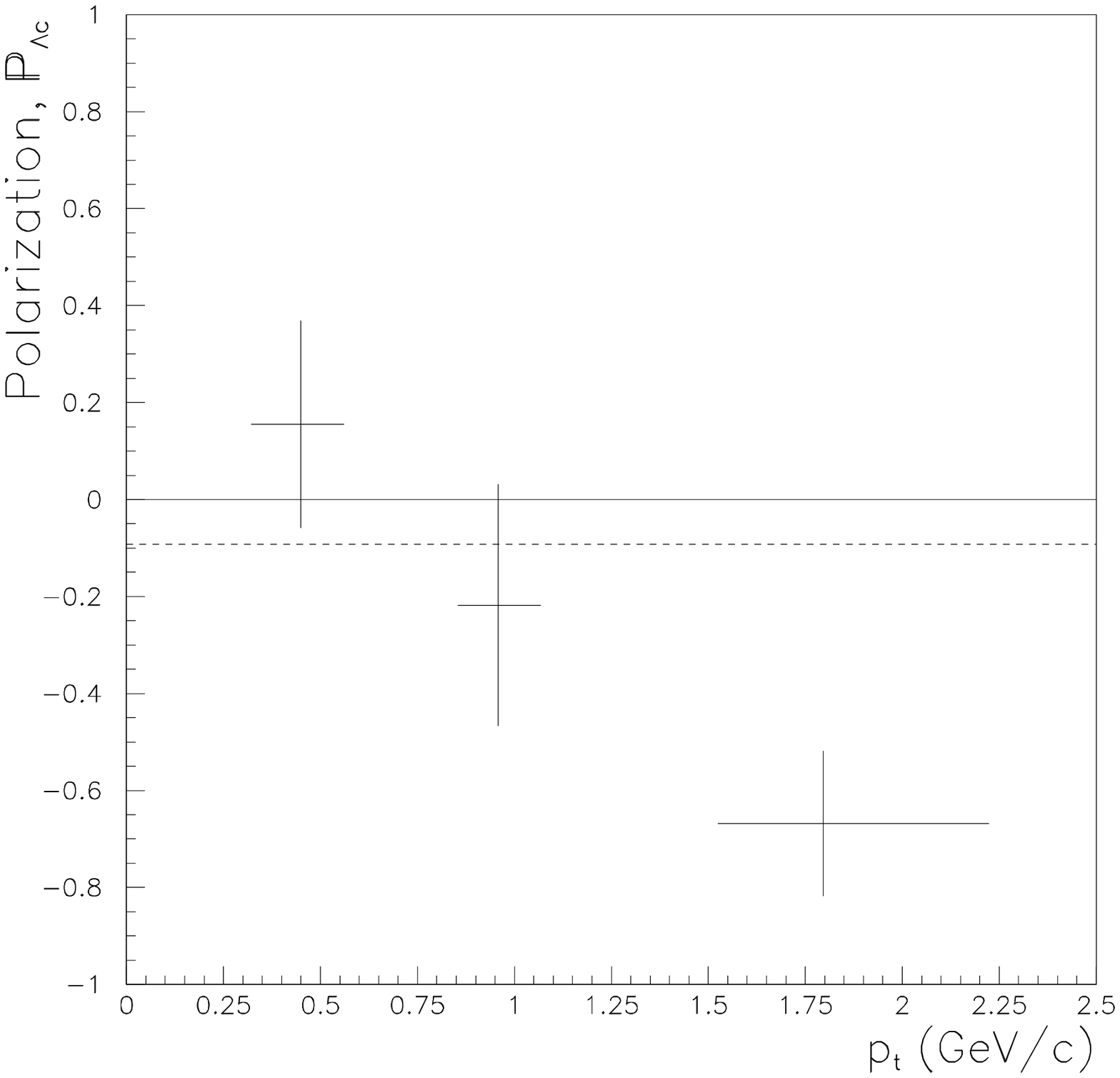}}
  \par\vspace{0pt}
 \end{minipage}
 \caption{{\bf Left:} Projections of fit (dashed lines) onto three mass
 pair and three angular variables (solid lines).
 Data lie in the range $2265<M(p\Km\pip)<2315$~MeV/c$^2$.
 {\bf Right:} $\Lc/\Lcbar$ Polarization from fit.}
 \end{figure}

The fit shown in Figure \ref{fig-fit_results} is seen to be good except
 for the low mass $\Km p$ region where an unmodelled enhancement is seen.
 Many $Y^{\ast}$ exist which could possibly account for this.
 Each $Y^{\ast}\pi$ channel added to our fit requires $\ge 4$ more
 parameters making it difficult, with our limited sample,
 to include more than one $Y^{\ast}$.  Adding $\Lambda(1600)\pi$,
 $\Sigma(1600)\pi$ or the tail of the $\Sigma(1405)\pi$ alone made no
 significant improvement.

Isobar fractions $f_R$ were computed by integrating over the
five dimensions of the fit $\vec x$:
\[ f_R =  \int\sum_{m,\lamp}\left|
          {\cal A}^{\ssc R}_{m,\lamp}\right|^2 d\vec x~{/}~
          \int\sum_{m,\lamp}\left|\sum_R
          {\cal A}^{\ssc R}_{m,\lamp}\right|^2 d\vec x    \]
Branching ratios with respect to $p\Km\pip$ are
 compared with earlier results in Table \ref{tab-brlc}.

 \begin{table}[h]
 \label{tab-brlc}
 \caption{First three columns are branching fractions relative to
  total $pK\pi$ mode (\% corrected for unseen decays).  The last four
  columns are resonant phases, described in the text, measured only
  by E791.}
 \begin{tabular*}{0.99\textwidth}{ c c c c c c c c }
 \bf Mode & \bf E791     & \bf NA32
          & \bf ISR
          & \tablehead{4}{c}{c}{E791 relative phases (degrees)}    \\
          &              & \citep{NA32_pkpi}
          & \citep{ISR_pkpi}
          & $\alpha_+$ & $\beta_+$ & $\alpha_-$ & $\beta_-$ \\ \hline
 $p\Kst(890)$
 & $29\pm 4\pm 3$
 & $35^{+6}_{-7}\pm 3$
 & $42\pm 24$
 & $ 58  \pm  28 $
 & $ 135 \pm  38 $
 & $ 198 \pm  24 $
 & $ 303 \pm  32 $
 \\
 $\Delta^{++}(1232)\Km$
 & $18\pm 3\pm 3$
 & $12^{+4}_{-5}\pm 5$
 & $40\pm 17$
 & $ 285 \pm 23  $
 & $ 280 \pm 23  $
 & $=\alpha_+$
 & $=\beta_+$
 \\
 $\Lambda(1520)\pip$
 & $15\pm 4\pm 2$
 & $9^{+4}_{-3}\pm 2$
 & --
 & $ 340 \pm 30  $
 & $ -3  \pm 32  $
 & $=\alpha_+$
 & $=\beta_+$
 \\
 $NR$
 & $55\pm 6\pm 4$
 & $56^{+7}_{-9}\pm 5$
 & --
 & $ 199 \pm 31  $
 & 0 (fixed)
 & $ 43  \pm 41  $
 & $ 65  \pm 21  $
 \\ \hline
 \end{tabular*}
 \end{table}

Good agreement is seen, but the significance of signals from NA32,
 where only mass projections were fit is overestimated.
NA32 errors are comparable to E791, but E791's sample is much
 larger.  This is because correlations among channels and relative
 phases were neglected in the NA32 analysis.

The $\Delta^{++}\Km$ mode is comparable to $p\Kstbar$ and clearly
 significant.

%% file: summary.tex
\section{Summary}
\vspace*{-15pt}

$\Lc$ production asymmetry in the range $-0.1<x_F<0.6$ and
 $p_T^2<8~(GeV/c)^2$ is constant at $\sim +0.15$
 favouring $\Lc$ over $\Lcbar$.  Models requiring a rising asymmetry
 toward negative $x_F$ are not ruled out however.
An amplitude analysis of the $\Lc$ decay shows the
 $\Lc\to\Delta^{++}\Km$ mode to be large indicating
 that the $W$ exchange amplitude is important.

\vspace*{-9pt}
\vspace*{-12pt}